\documentclass{jpsj3}
\usepackage[dvipdfmx]{graphicx}
\usepackage{txfonts}
\usepackage{multirow}
\usepackage{times}
\usepackage{here} 
\usepackage{bm}
\usepackage{braket}
\usepackage{setspace}
\usepackage{amsmath}
\usepackage{amsfonts}
\usepackage{lscape}
\usepackage{mathrsfs}
\usepackage{ascmac}
\usepackage{cases}
\usepackage{here}

\title{Paramagnetic effects of $j$-electron superconductivity and application to UTe$_2$ }

\author{Kozo Hiranuma$^1$, Satoshi Fujimoto$^1$}
\inst{$^1$Department of Materials Engineering Science, Osaka University, Toyonaka, Osaka 560-8531,  Japan}

\abst{The putative spin-triplet superconductor UTe$_2$ 
has an Ising-like magnetic anisotropy, indicating the Cooper pair spins parallel to the easy-axis.
However, the upper critical field for the hard-axis direction is much larger than that for the easy-axis,
implying the strong suppression of the Pauli depairing effect even for the hard-axis direction. 
This observation is consistent with the recent NMR measurements which show the tiny decrease of the Knight shift below the
transition temperature $T_c$ for hard-axis magnetic fields.
We clarify that these behaviors are naturally understood as a result of multi-orbital $f$-electron bands with total angular momentum $j=5/2$.
Taking into account spin-orbit couplings for an orthorhombic structure, we demonstrate that for hard-axis magnetic fields,
the decrease of the spin susceptibility below $T_c$ and the Pauli depairing effect are strongly suppressed.}

\begin{document}
\maketitle

\section{Introduction}
Recently, the U-based heavy-fermion system UTe$_2$ was newly discovered as a candidate material of spin-triplet superconductivity (SC).{\cite{1,2}} The crystal structure of this material is orthorhombic (the point group $D_{2h}$) and has a strongly Ising-like magnetic anisotropy along the $a$-axis.{\cite{3,4}} One of the most interesting properties of UTe$_2$ is the anomalously large upper critical field $H_{c2}$.{\cite{1,2,butch1,knafo}} The $H_{c2}$ is highly anisotropic and well above the Pauli limit estimated from the BCS theory for all the crystal axis directions. In addition, $H_{c2}~{\perp}~a$-axis greatly exceed $H_{c2}~{\parallel}~a$-axis, especially $H_{c2}~{\parallel}~b$-axis, which is noticeably larger. These properties of $H_{c2}$ cannot be explained by spin-singlet pairings, and are strong signatures of  a spin-triplet pairing state.\par
These features, i.e. the orthorhombic crystal structure, a strong Ising-like magnetic anisotropy, and anomalously large and anisotropic $H_{c2}$, 
 are quite similar to the properties of the U-based ferromagnetic superconductors (FM-SCs),{\cite{5,6}} UGe$_2$,\cite{7} UCoGe\cite{8} and URhGe,\cite{9}
in which superconductivity coexists with ferromagnetic orders.
However, an important difference is that UTe$_2$ is paramagnetic above $T_c$, and there is no experimental evidence of the coexistence of superconductivity and ferromagnetism. In the case of the FM-SCs, the anomalously large and strongly anisotropic upper critical fields can be understood as follows.\cite{10,11} Since the pairing glue is the longitudinal ferromagnetic spin fluctuation along the magnetic easy axis,
the pairing interaction is weakened for magnetic fields parallel to the easy axis, which suppress the longitudinal ferromagnetic spin
fluctuation. 
Therefore, $H_{c2}~{\parallel}$ the easy-axis is substantially small compared to the other directions. 
On the other hand, the longitudinal ferromagnetic spin fluctuation favors spin-triplet Cooper pairs with spins parallel to the easy axis.
This implies that the Pauli depairing effect may exist for magnetic fields perpendicular to the easy axis. 
The formulae for the free energy of the normal and SC states are $F_{N}=E_0-\frac{1}{2}\chi_{N}H^2$ and $F_{SC}=E_0-\frac{1}{2}\chi_{SC}H^2-\frac{1}{2}N(0)\Delta^2$, which are equal at the Pauli limitting field.
Hence, the Pauli limiting field is given by.
\begin{eqnarray}
H_{p}=\Delta\sqrt{\frac{N(0)}{\delta\chi_{spin}}}
\label{Pauli limit by BCS}
\end{eqnarray}
Here, $\Delta$ is the SC gap, $N(0)$ is the density of states on the Fermi surface, and $\delta\chi_{spin}=\chi_{N}-\chi_{SC}$ is the decrease in the spin susceptibility due to the superconducting transition. 
In the case of the FM-SCs, $\delta\chi_{spin}$ is very small provided that the exchange splitting of the Fermi surface is sufficiently larger than the SC gap.\cite{10} Hence, the Pauli depairing effect for magnetic fields perpendicular to the easy axis is suppressed.  In fact, the small $\delta\chi_{spin}$ of UCoGe was experimentally verified via the NMR Knight shift measurements.\cite{12} Thus, the anomalous behaviors of $H_{c2}$ in the FM-SCs are explained by ferromagnetism.
However, the same explanation is not applicable to UTe$_2$, because it is paramagnetic.\par

On the other hand, according to the recent NMR Knight shift experiment for UTe$_2$,\cite{13} the decrease of $\chi_{spin}$ for magnetic fields parallel to 
hard axis directions
is merely $\sim 8\%$, which allows us to understand the absence of the Pauli depairing effect for magnetic fields perpendicular to the easy axis. 
Then, an important question is why $\delta\chi_{spin}$ becomes so small even in the absence of the exchange splitting due to ferromagnetism.

In this paper, we propose a scenario that 
the small $\delta\chi_{spin}$ of UTe$_2$ is caused by band splitting due to the spin-orbit (SO) coupling of $f$-electrons with the total angular momentum $j=5/2$. We apply a microscopic analysis to a multi-orbital model of $j=5/2$ electrons with the SO coupling of an orthorhombic structure to derive formulae of the spin susceptibility and perform numerical calculations. Here, we mainly discuss $\delta\chi_{spin}$ for hard-axis magnetic fields,
which is directly related to the recent NMR measurements.\cite{13} 
For comparison, the case of the easy axis ($a$-axis) and the case of spin-singlet pairings are also briefly described in Appendix.

The organization of this paper is as follows.
In the section 2, we present a model and the formulae of the spin susceptibility. 
In the section 3,  a discussion of candidate pairing states is given.
The numerical results are presented in  the section 4.
In Appendix, we present technical details, and also, some numerical results for the case with the easy-axis magnetic field and the case of spin-singlet pairings for the sake of comparison. 

\section{Model and formulae of spin susceptibility}
In this section, we describe the model and method used in calculations. First, we notice that for UTe$_2$, the conduction band is composed of $f$-orbital electrons with the total angular momentum $j=5/2$ hybridized with $d$-orbital electrons of $U$ and $p$-orbital electron of Te, as
clarified by the band calculation using DFT and GGA+U.\cite{2,14,band} 
To simplify the analysis, we consider the model in which the electron bands are formed only by  $j=5/2$ electrons, and
take into account the SO interaction acting on $j=5/2$ in the orthorhombic structure.
Because of the $D_{2h}$ point group symmetry and the strong Ising anisotropy of UTe$2$,
we postulate the following form of the SO interaction,
\begin{eqnarray}
\hat{\mathcal{H}}_{SO}=\sum_{\mu=a,b,c}\lambda_\mu\hat{J}_\mu^2k^2_\mu\quad(\lambda_\mu\leq0).
\label{eq:SO}
\end{eqnarray}
Here, $\lambda_\mu$ is the coupling constant of the SO interaction, and $\hat{J}_\mu$ is the total angular momentum operator for $j=5/2$. 
This spin-orbit interaction is due to the hybridization of  $J$-multiples split by crystalline fields between neighboring sites, and derived from the standard $\boldsymbol{k}\cdot\boldsymbol{p}$ perturbation theory.\cite{luttinger}
Since the system has the strong Ising anisotropy in the $a$-axis direction, we can properly assume that $\lambda_a \neq 0$, 
and $\lambda_b=\lambda_c=0$, which simplifies the analysis substantially. 
We also choose the spin quantization axis parallel to the $a$-axis.
Thus, $\hat{J}_a$ is diagonal.
Then, the second-quantized Hamiltonian for the normal state is written in the following form,
\begin{eqnarray}
\hat{\mathcal{H}}=\sum_{\bm{k},\alpha}\left[\varepsilon_{k}+\lambda_a(\hat{J}_a)^2_{\alpha,\alpha}k^2_a-\mu\right]\hat{c}_{\bm{k}\alpha}^{\dagger}\hat{c}_{\bm{k}\alpha},
\end{eqnarray}
where $\varepsilon_{k}$ is the band dispersion in the case without the SO interaction, $\mu$ is a chemical potential, and 
$\hat{c}_{\bm{k}\alpha}$ ($\hat{c}_{\bm{k}\alpha}^{\dagger}$) is an annihilation (creation) operator of an electron with the momentum $\bm{k}$ and
the spin projection $j_a=\alpha$.
The SO interaction (\ref{eq:SO}) splits the $j=5/2$ band into three Kramers pair bands with $j_a=\pm5/2,\ \pm3/2\ \mathrm{and}\ \pm1/2$. We can plausibly assume that the SO splitting is much larger than the SC gap for $f$-electron systems, and that
Cooper pairs between bands with different $|j_a|$ are energetically unstable, and only intra-band Cooper pairs are
formed, as shown below.
\begin{eqnarray}
\Delta_{j_a,j_a'}=
\begin{cases}
0 & |j_a|\ne|j_a'|\\
finite & |j_a|=|j_a'|
\end{cases}
\end{eqnarray}
In this way, we only have to deal with three $2\times2$ Gor'kov equations instead of a $6\times6$ equation for the derivation of the normal and anomalous Green functions $\hat{G}_{j_a}$ and $\hat{F}_{j_a}$ (for details of the derivation, see Appendix).
We derive the formulae of spin susceptibility by using the linear response theory (see Appendix), which finally yield the following formulae 
for the longitudinal spin susceptibility $\chi_{aa}$ and the transverse spin susceptibility $\chi_{bb}$, 
\begin{eqnarray}
\label{chiaa}
\chi_{aa}&=&\sum_{|j_a|=\frac{1}{2},\frac{3}{2},\frac{5}{2}}\chi_{aa}^{\{|j_a|\}}=\chi_{aa}^{\{\frac{1}{2}\}}+\chi_{aa}^{\{\frac{3}{2}\}}+\chi_{aa}^{\{\frac{5}{2}\}}\\
\label{chibb}
\chi_{bb}&=&\chi_{bb}^{\{\frac{1}{2}\}}+\sum_{\{\alpha,\beta\}}\chi_{bb}^{\{\alpha,\beta\}}\nonumber\\
&=&\chi_{bb}^{\{\frac{1}{2}\}}+\chi^{\{\frac{3}{2},\frac{1}{2}\}}_{bb}+\chi^{\{-\frac{1}{2},-\frac{3}{2}\}}_{bb}+\chi^{\{\frac{5}{2},\frac{3}{2}\}}_{bb}+\chi^{\{-\frac{3}{2},-\frac{5}{2}\}}_{bb}
\end{eqnarray}
where $\chi_{aa}^{\{|j_a|\}}$ ($j_a= \frac{1}{2}, \frac{3}{2}, \frac{5}{2}$) is the longitudinal spin susceptibility for the band with $\pm j_a$,
$\chi_{bb}^{\{\frac{1}{2}\}}$ is the transverse spin susceptibility for the band with $j_a=\pm 1/2$, 
and $\chi_{bb}^{\{\alpha,\beta\}}$ is the transverse spin susceptibility for the transition between the bands with $j_a=\alpha$ and $\beta$, where
$\{\alpha,\beta\}=\{\frac{3}{2},\frac{1}{2}\},\{-\frac{1}{2},-\frac{3}{2}\},\{\frac{5}{2},\frac{3}{2}\},\{-\frac{3}{2},-\frac{5}{2}\}$.
In $\chi_{aa}$, all terms are intra-band contributions, while in $\chi_{bb}$, all the terms other than $\chi_{bb}^{\{\frac{1}{2}\}}$ are
 inter-band contributions. See Appendix3 for detailed formulae.

\section{Candidate pairing states}
In this section, we discuss possible pairing states, and the $d$-vector that determines the gap structure of the spin-triplet superconducting state. The basis functions of the $d$-vector are classified according to the irreducible representations (IRs) of the point group symmetry of the system. UTe$_2$ has $D_{2h}$ crystal symmetry, and its IRs in the case of odd-parity pairing states are $A_u$, $B_{1u}$, $B_{2u}$ and $B_{3u}$ representations. The $d$-vectors of each IRs are as follows.\cite{14}
\begin{eqnarray}
\bm{d}^{A_u}&=&\gamma^{A_u}_ak_a\hat{a}+\gamma^{A_u}_bk_b\hat{b}+\gamma^{A_u}_ck_c\hat{c}\\
\bm{d}^{B_{1u}}&=&\gamma^{B_{1u}}_ak_a\hat{a}+\gamma^{B_{1u}}_bk_b\hat{b}\\
\bm{d}^{B_{2u}}&=&\gamma^{B_{2u}}_ak_a\hat{a}+\gamma^{B_{2u}}_ck_c\hat{c}\\
\bm{d}^{B_{3u}}&=&\gamma^{B_{3u}}_bk_b\hat{b}+\gamma^{B_{3u}}_ck_c\hat{c}
\end{eqnarray}
Here, $\gamma^{IR}_\mu\ (IR=A_u,B_{1u},B_{2u},B_{3u}\ \mathrm{and}\ \mu=a,b,c)$ are complex constants.
Since the system has the strong Ising-like anisotropy along the $a$-axis, it is plausible to assume that
the $a$-component of each $d$-vector is suppressed, at least, in the absence of applied magnetic fields.
We, first, consider the case of the unitary state with time-reversal symmetry.
Among the above four representations, the $A_u$ state is full gap, though,
if the $a$-components of the $d$-vectors are zero because of the above reason, it has point nodes.
The $B_{3u}$ state also has point nodes.
The other two representations have line nodes.
Thus, the $A_u$ and $B_{3u}$ states are energetically more stable, and possible candidates. On the other hand, in the case with applied magnetic fields which lower the symmetry of the system, as in the case of NMR experiments,\cite{13}
the admixture of different IRs is allowed. Furthermore, it is possible 
that non-unitary states with broken time-reversal symmetry are realized in the admixture states. In the case with a strong magnetic field applied along the $\mu$-axis, Cooper pairs in a non-unitary state are expected to carry the magnetization in the $\mu$-axis direction. This condition is expressed by the following equation,
\begin{eqnarray}
\label{qb}
\int_{\mathrm{FS}}q_\mu{d^2k}\ne0,
\end{eqnarray}
where $q_\mu$ represents the $\mu$ direction component of $\bm{q}=i\bm{d}\times\bm{d}^*$, and the integration is performed on the Fermi surface. 
We consider the case that a magnetic field is parallel to the $b$-axis.
Then, the admixture state of the $B_{1u}+B_{3u}$ representation, or that of the $A_{u}+B_{2u}$ representation is possible.
For instance, we, here, focus on the $B_{1u}+B_{3u}$ state, because qualitative behaviors of spin susceptibility 
are similar between these states, as checked by numerical calculations.
 The followings are two candidates of $d$-vectors belonging to the $B_{1u}+B_{3u}$ representation and satisfying the condition Eq.(\ref{qb}) with $\mu=b$,
\begin{eqnarray}
\bm{d}^{B_{1u}+B_{3u}}_1&=&k_b\hat{a}+k_c\hat{b}+ik_b\hat{c}\\
\bm{d}^{B_{1u}+B_{3u}}_2&=&ik_b\hat{a}+(k_c+ik_a)\hat{b}+k_b\hat{c}
\end{eqnarray}
Here, for simplicity, we set all coefficients to 1 or $i$. In the case of $\bm{d}^{B_{1u}+B_{3u}}_1$, there is a line node at the crossing line of the $k_a$-$k_b$ plane and the Fermi surface. In the case of $\bm{d}^{B_{1u}+B_{3u}}_2$, there are point nodes at the crossing points of the $k_b$-axis and the Fermi surface, and these are Weyl point nodes, which carry monopole charges and are the source and the drain of the Berry curvature
in the momentum space. That is, the non-unitary state with $\bm{d}^{B_{1u}+B_{3u}}_2$ is the Weyl superconducting state.\cite{fujimoto}
The $d$-vectors considered in this paper are summarized in Table I.
In the next section, we present numerical results of spin susceptibility calculated for these pairing states.
\begin{table}[h]
\caption{$d$-vector used in numerical calculations}
\begin{center}
\begin{tabular}{c|ccccc}
\hline\hline
magnetic field&$\bm{d}$&$\bm{q}=i\bm{d}\times\bm{d}^*$&Gap structure&IR&\\
\hline
\multirow{2}{*}{$\bm{H}=0$ or $\bm{H}\parallel{b}$-axis}&$(k_a,k_b,k_c)$&$0$&Full gap or point node&$A_u$&\multirow{2}{*}{Unitary}\\
&$(0,k_c,k_b)$&$0$&Point node&$B_{3u}$&\\
\hline
\multirow{2}{*}{$\bm{H}\parallel{b}$-axis}&$(k_b,k_c,ik_b)$&$(2k_bk_c,2k_b^2,0)$&Line node&$B_{1u}+B_{3u}$&\multirow{2}{*}{Non-unitary}\\
&$(ik_b,k_c+ik_a,k_b)$&$(-2k_ak_b,2k_b^2,-2k_bk_c)$&Point nodes, Weyl&$B_{1u}+B_{3u}$&\\
\hline\hline
\end{tabular}
\end{center}
\label{d-vectors}
\end{table}

\section{Results of numerical calculations of spin susceptibillity}
In this section, we present the results of numerical calculations of spin susceptibility for magnetic field parallel to the $b$-axis, $\chi_{bb}$, in the cases of the $d$-vectors shown in Table \ref{d-vectors}.  In the calculations, we set the chemical potential $\mu=1$ and the SC transition temperature $T_c=0.01$, and vary the magnitude of the SO coupling $|\lambda_a|$ from 0 to 0.25. 
We, first, show the results for the unitary state ($A_u$ and $B_{3u}$ representations) in Fig. 1. The decreases in $\chi_{bb}$ below $T_c$ are approximately $7\sim9\%$ in the case of $\bm{d}^{A_u}=(k_a,k_b,k_c)$, and $11\sim13\%$ in the case of $\bm{d}^{B_{3u}}=(0,k_c,k_b)$. 
In the insets of Figure 1, the contributions of each term in Eq. (\ref{chibb}) for $|\lambda_a|=0.10$ are depicted.
As seen in these insets, 
only the intra-band contribution $\chi_{bb}^{\{\frac{1}{2}\}}$ (light blue line) decreases, while the inter-band contributions $\chi_{bb}^{\{\alpha,\beta\}}$( dark and light yellow lines) do not change at all.
\begin{figure}[H]
\begin{center}
\includegraphics[clip,width=13.5cm]{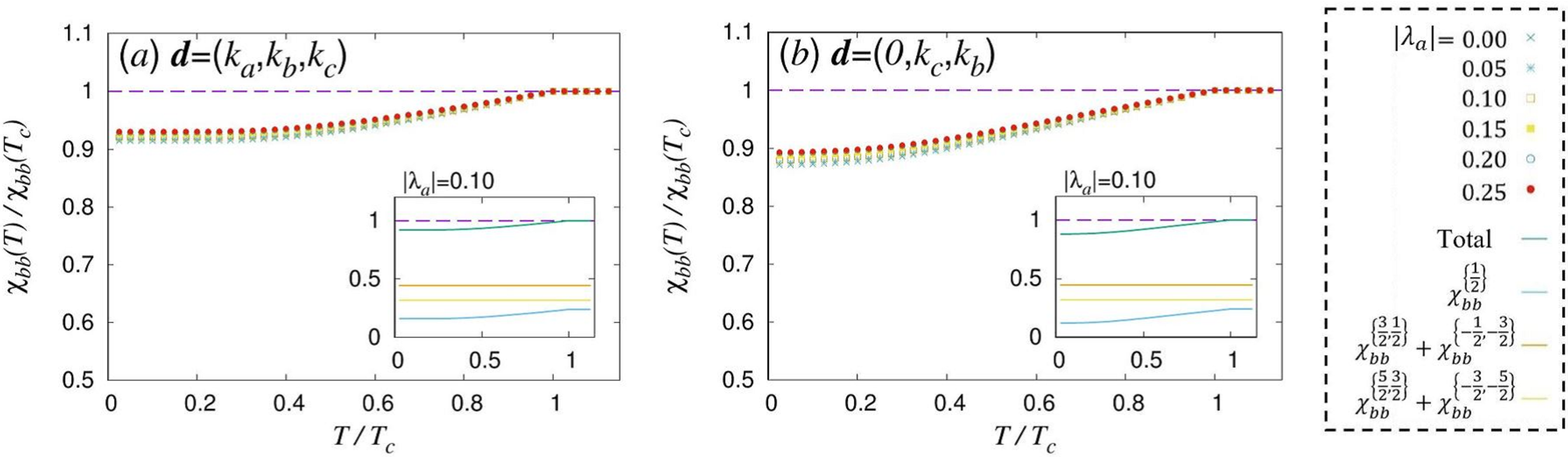}
\end{center}
\caption{(Color)The temperature dependence of $\chi_{bb}$ for the unitary state. (a)$A_u$ : $\bm{d}=(k_a,k_b,k_c)$. (b)$B_{3u}$ : $\bm{d}=(0,k_c,k_b)$.)}
\end{figure}
Next, we show the results for the non-unitary state ($B_{1u}+B_{3u}$ representation) in Fig. 2. The decreases in $\chi_{bb}$ are $10\sim12\%$ in the case of $\bm{d}^{B_{1u}+B_{3u}}_1=(k_b,k_c,ik_b)$, and $13\sim15\%$ in the case of $\bm{d}^{B_{1u}+B_{3u}}_2=(ik_b,k_c+ik_a,k_b)$. Here, as in the case of the unitary, in insets of both cases, only the intra-band contribution $\chi_{bb}^{\{\frac{1}{2}\}}$ (light blue line) decreases, while the inter-band contributions $\chi_{bb}^{\{\alpha,\beta\}}$ (dark and light yellow lines) do not change at all.
These results confirm that, in common with both the unitary and non-unitary cases, the decreases are significantly suppressed, and only the intra-band term with $j_a=\pm1/2$ contributes to the decrease. In other words, the decreases in $\chi_{bb}$ are considered to be suppressed by the band splitting due to the SO coupling. 
Our results provide a clear explanation of the small decrease of the Knight shift, i.e.  $\sim 8\%$ of the spin part, observed in the recent NMR experiment.\cite{13}\par
\begin{figure}[h]
\begin{center}
\includegraphics[clip,width=13.5cm]{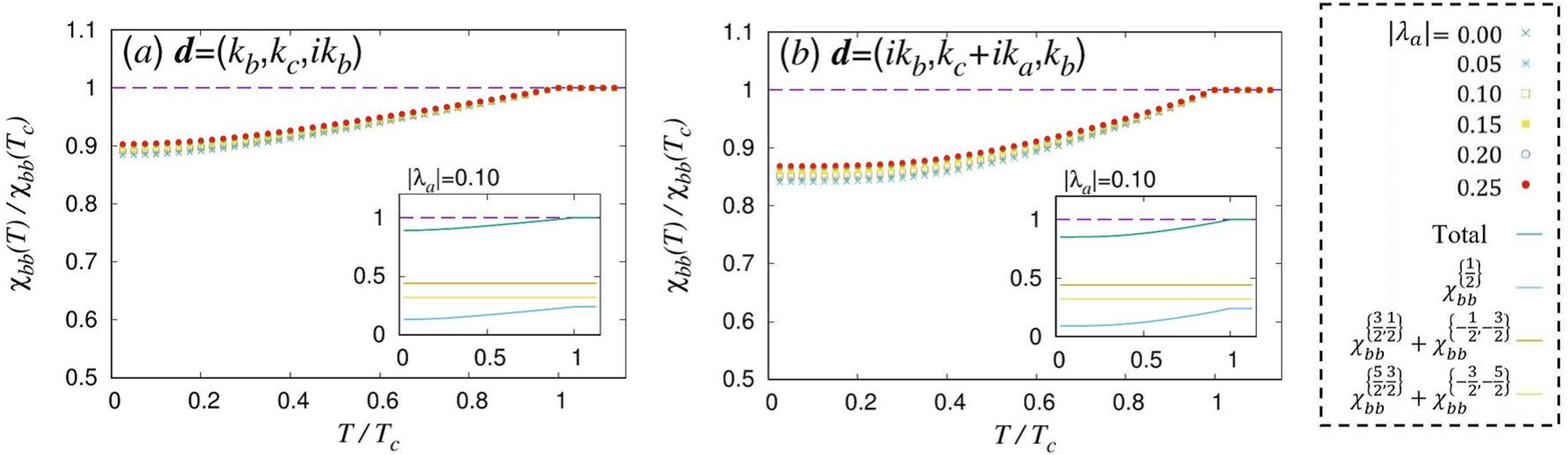}
\end{center}
\caption{(Color)The temperature dependence of $\chi_{bb}$ for the non-unitary state. (a)$B_{1u}+B_{3u}$ : $\bm{d}=(k_b,k_c,ik_b)$. (b)$B_{1u}+B_{3u}$ : $\bm{d}=(ik_b,k_c+ik_a,k_b)$.}
\end{figure}
It is noted that, as seen in Figs. 1and 2, even in the case of $\lambda_a=0$, the decrease of $\chi_{bb}$ is very small.
This is because that in our model Cooper pairs are formed only within the Kramers degenerate band, and inter-band pairs formed between
 electrons with different $|j_a|$ are completely neglected. This approximation is valid only when the SO splitting is much larger than
 the SC gap. Since there is no inter-band pairings even in the limit of vanishing SO splitting in our model, the decrease of the transverse susceptibility
 for the transition from $j_a$ to $j_a\pm 1$ is strongly suppressed.  We stress that for realistic situations, $\lambda_a k_F \gg \Delta$ is satisfied. 

We also note that the strong suppression of $\delta \chi_{spin}$ due to the SO interaction of $j$-electrons occurs only for the transverse susceptibility.
The longitudinal spin susceptibility $\chi_{aa}$ is not affected by the SO splitting, since it consists of only intra-band contributions.
For comparison, we present calculated results of $\chi_{aa}$ in Appendix A.4.

Before closing this section, we remark on the case of spin-singlet pairings. The strong suppression of the Pauli depairing effect
for hard-axis magnetic fields in $j$-electron superconductivity takes place also in spin-singlet pairing states, as long as the SO splitting is sufficiently
larger than the SC gap. However, as mentioned above, the longitudinal spin susceptibility decreases toward zero in this case, and the
Pauli limiting field for easy-axis magnetic fields exhibits a conventional behavior of spin-singlet pairing states. (see Appendix A.5)

\section{Conclusion}
In this paper, in order to elucidate the origin of small $\delta\chi_{spin}$ for magnetic fields parallel to the $b$-axis observed in the NMR Knight shift measurement of UTe$_2$,\cite{12} we numerically calculated the transverse spin susceptibility $\chi_{bb}$ using the $j=5/2$ multi-orbital model. 
As a result, the decrease in $\chi_{bb}$ is sufficiently small for each $d$-vector candidate, implying that the band splitting due to the SO interaction can explain the results of the NMR Knight shift experiment\cite{13} and the anomalously large upper critical field of UTe$_2$ for magnetic fields in the hard-axis direction.\cite{1,2,butch1}\par
Finally, we note that the results of the transverse spin susceptibility $\chi_{bb}$ mentioned above are not changed qualitatively even if we use
any other pairing states. Thus, we can conclude that the Pauli depairing effect for hard axis magnetic fields is generally suppressed for $j$-electron superconductivity. From this point of view, we expect that this mechanism plays an important role also for the U-based FM-SCs, UGe$_2$, UCoGe, and URhGe, since the energy scales of the SO splitting are much larger than those of the exchange splitting in these systems.  

\begin{acknowledgment}
This work was supported by the Grant-in-Aids for Scientific
Research from MEXT of Japan [Grants No. 17K05517], and KAKENHI on Innovative Areas ``Topological Materials Science'' [No.~JP15H05852]  and "J-Physics" [No.~JP18H04318], and JST CREST Grant Number JPMJCR19T5, Japan. 
\end{acknowledgment}

\appendix

\section{Derivation of spin susceptibility formulae}
Assuming that Cooper pairs are formed within the intra-bands, we can write down the Gor'kov equations for the multi-orbital model in the following forms,
\begin{eqnarray}
\left\{i\omega_m-\varepsilon_{k,j_a}\right\}\hat{G}_{j_a}-{\hat{\Delta}}_{j_a}\hat{F}_{j_a}^{\dagger}&=&\hat{\sigma}_0, \\
\left\{i\omega_m+\varepsilon_{k,j_a}\right\}\hat{F}_{j_a}^{\dagger}-{\hat{\Delta}_{j_a}}^{\dagger}\hat{G}_{j_a}&=&0.
\end{eqnarray}
Here, the Matsubara frequency $\omega_m=(2m+1)\pi$, $\varepsilon_{k,j_a}=\varepsilon_{k}+{\lambda}_aj_a^2k^2_a-\mu$, $\hat{\sigma}_0$ is $2\times2$ unit matrix. The normal and anomalous Green functions $\hat{G}_{j_a}$ and $\hat{F}_{j_a}$ and the anomalous self-energy part $\hat{\Delta}_{j_a}$ are $2\times2$ matrices of the space spanned by the states with $\pm{j_a}$. The vector of the three Pauli matrices are defined as follows.
\begin{eqnarray}
\bm{\hat{\sigma}}=\left(\hat{\sigma}_a,\hat{\sigma}_b,\hat{\sigma}_c\right)=\left(
\begin{bmatrix}
1&0\\
0&-1\\
\end{bmatrix}
,
\begin{bmatrix}
0&1\\
1&0\\
\end{bmatrix}
,
\begin{bmatrix}
0&-i\\
i&0\\
\end{bmatrix}
\right).
\end{eqnarray}
\noindent

Then, $\hat{G}_{j_a}$, $\hat{F}_{j_a}$ and $\hat{\Delta}_{j_a}$ are represented in the following way in terms of $\hat{\sigma}$ and $\bm{\hat{\sigma}}$.
\begin{eqnarray}
\hat{G}_{j_a}&=&
\begin{bmatrix}
G_{j_a,j_a}&G_{j_a,-j_a}\\
G_{-j_a,j_a}&G_{-j_a,-j_a}\\
\end{bmatrix}
=G^{j_a}_0\hat{\sigma}_0+\bm{G}^{j_a}\cdot\bm{\hat{\sigma}}=
\begin{bmatrix}
G^{j_a}_0+G^{j_a}_a&G^{j_a}_b-iG^{j_a}_c\\
G^{j_a}_b+iG^{j_a}_c&G^{j_a}_0-G^{j_a}_a\\
\end{bmatrix},
\\
\hat{F}_{j_a}&=&
\begin{bmatrix}
F_{j_a,j_a}&F_{j_a,-j_a}\\
F_{-j_a,j_a}&F_{-j_a,-j_a}\\
\end{bmatrix}
=\left(F^{j_a}_0\hat{\sigma}_0+\bm{F}^{j_a}\cdot\bm{\hat{\sigma}}\right)i\hat{\sigma}_c=
\begin{bmatrix}
-F^{j_a}_b+iF^{j_a}_c&F^{j_a}_0+F^{j_a}_a\\
-F^{j_a}_0+F^{j_a}_a&F^{j_a}_b+iF^{j_a}_c\\
\end{bmatrix},
\\
\hat{\Delta}_{j_a}&=&
\begin{bmatrix}
\Delta_{j_a,j_a}&\Delta_{j_a,-j_a}\\
\Delta_{-j_a,j_a}&\Delta_{-j_a,-j_a}\\
\end{bmatrix}
=\left(d_0\hat{\sigma}_0+\bm{d}\cdot\bm{\hat{\sigma}}\right)i\hat{\sigma}_c=
\begin{bmatrix}
-d_b+id_c&d_0+d_a\\
-d_0+d_a&d_b+id_c\\
\end{bmatrix}.
\end{eqnarray}

\subsection{Linear response theory}
The magnetic moment operator in the $\mu$ direction is written in the following form,
\begin{eqnarray}
\hat{M}_{\bm{q}\mu}(\tau)=-g\mu_B\sum_{\bm{k},\alpha,\beta}\hat{c}_{\bm{k}-\bm{q},\alpha}^\dagger(\tau)(\hat{J}_{\mu})_{\alpha\beta}\hat{c}_{\bm{k},\beta}(\tau).\quad(\mu=a,b,c)
\end{eqnarray}
Here, $g$ is a $g$-factor, $\mu_{\rm B}$ is the Bohr magneton, and  $\hat{J}_\mu$ are the total angular momentum operators with $j=5/2$ and are expressed as follows,
\begin{eqnarray}
\label{ja}
\hat{J}_a&=&\frac{1}{2}
\begin{bmatrix}
5&0&0&0&0&0\\
0&3&0&0&0&0\\
0&0&1&0&0&0\\
0&0&0&-1&0&0\\
0&0&0&0&-3&0\\
0&0&0&0&0&-5\\
\end{bmatrix},\\
\label{jb}
\hat{J}_b&=&\frac{1}{2}
\begin{bmatrix}
0&\sqrt{5}&0&0&0&0\\
\sqrt{5}&0&2\sqrt{2}&0&0&0\\
0&2\sqrt{2}&0&3&0&0\\
0&0&3&0&2\sqrt{2}&0\\
0&0&0&2\sqrt{2}&0&\sqrt{5}\\
0&0&0&0&\sqrt{5}&0\\
\end{bmatrix},\\
\label{jc}
\hat{J}_c&=&\frac{i}{2}
\begin{bmatrix}
0&-\sqrt{5}&0&0&0&0\\
\sqrt{5}&0&-2\sqrt{2}&0&0&0\\
0&2\sqrt{2}&0&-3&0&0\\
0&0&3&0&-2\sqrt{2}&0\\
0&0&0&2\sqrt{2}&0&-\sqrt{5}\\
0&0&0&0&\sqrt{5}&0\\
\end{bmatrix}.
\end{eqnarray}
According to the linear response theory, the correlation function of the magnetic moment $\hat{M}_{\bm{q}\nu}(\tau)$  is written in the following form,
\begin{eqnarray}
\Phi_{\mu\nu}(\bm{q},i\omega_n)&=&\int_0^\beta\Braket{T_\tau\left[\hat{M}_{\bm{q}\mu}(\tau)\hat{M}_{-\bm{q}\nu}(0)\right]}e^{i\omega_n\tau}d\tau\nonumber\\
&=&g^2\mu_B^2\sum_{\substack{\bm{k},\bm{k}' \\ \alpha,\beta,\gamma,\delta}}(\hat{J}_{\mu})_{\alpha\beta}(\hat{J}_{\nu})_{\gamma\delta}\int_0^\beta\Braket{T_\tau\left[\hat{c}_{\bm{k}-\bm{q},\alpha}^\dagger(\tau)\hat{c}_{\bm{k},\beta}(\tau)\hat{c}_{\bm{k}'+\bm{q},\gamma}^\dagger(0)\hat{c}_{\bm{k}',\delta}(0)\right]}e^{i\omega_n\tau}d\tau. \nonumber\\
\label{A1}
\end{eqnarray}
Applying the Bloch-De Dominisis theorem to (\ref{A1}), and introducing the following normal and anomalous Green functions,
\begin{eqnarray}
&&G_{\alpha,\beta}(\bm{k},\tau-\tau')=-\Braket{T_\tau\left[\hat{c}_{\bm{k},\alpha}(\tau)\hat{c}_{\bm{k},\beta}^\dagger(\tau')\right]}=\frac{1}{\beta}\sum_me^{-i\omega_m(\tau-\tau')}G_{\alpha,\beta}(\bm{k},i\omega_m),\\
&&F_{\alpha,\beta}^{\dagger}(\bm{k},\tau-\tau')=-\Braket{T_\tau\left[\hat{c}_{-\bm{k},\alpha}^{\dagger}(\tau)\hat{c}_{\bm{k},\beta}^{\dagger}(\tau')\right]}=\frac{1}{\beta}\sum_me^{-i\omega_m(\tau-\tau')}F^{\dagger}_{\alpha,\beta}(\bm{k},i\omega_m),\\
&&F_{\alpha,\beta}(\bm{k},\tau-\tau')=-\Braket{T_\tau\left[\hat{c}_{\bm{k},\alpha}^{\quad}(\tau)\hat{c}_{-\bm{k},\beta}^{\quad}(\tau')\right]}=\frac{1}{\beta}\sum_me^{-i\omega_m(\tau-\tau')}F_{\alpha,\beta}(\bm{k},i\omega_m),
\end{eqnarray}
we rewrite (\ref{A1}) into the following form,
\begin{eqnarray}
\label{a4}
\Phi_{\mu\nu}(\bm{q},i\omega_n)&=&-\frac{g^2{\mu_B}^2}{\beta}\sum_{\substack{\bm{k},m,\alpha \\ \beta,\gamma,\delta}}(\hat{J}_{\mu})_{\alpha\beta}(\hat{J}_{\nu})_{\gamma\delta}F_{\alpha,\gamma}^{\dagger}(-\bm{k}+\bm{q},i\omega_m-i\omega_n)F_{\beta,\delta}(\bm{k},-i\omega_m)\nonumber\\
&&-\frac{g^2{\mu_B}^2}{\beta}\sum_{\substack{\bm{k},m,\alpha \\ \beta,\gamma,\delta}}(\hat{J}_{\mu})_{\alpha\beta}(\hat{J}_{\nu})_{\gamma\delta}G_{\delta,\alpha}(\bm{k}-\bm{q},i\omega_m+i\omega_n)G_{\beta,\gamma}(\bm{k},i\omega_m).
\end{eqnarray}
The spin susceptibility $\chi_{\mu\nu}$ is obtained by analytically continuing the correlation function $\Phi_{\mu\nu}(\bm{q},i\omega_n)$ and taking $\bm{q}$-limit.
\begin{eqnarray}
\label{a5}
\Phi_{\mu\nu}(\bm{q},i\omega_n)\xrightarrow{i\omega_n{\rightarrow}\omega+i\delta}\chi_{\mu\nu}(\bm{q},\omega+i\delta)\xrightarrow{\bm{q}-\mathrm{limit},\,\delta\rightarrow0}\chi_{\mu\nu}
\end{eqnarray}
Here, $\bm{q}$-limit means to take the limits of $\omega\rightarrow0$ before $\bm{q}\rightarrow0$.\par
We show the precise expressions of $\Phi_{\mu\nu}$ in the cases of $\mu=\nu=a$ and $\mu=\nu=b$ in the following. The formulae shown below can be divided into several terms, and the subscripts $\{\alpha\}$ and $\{\alpha,\beta\}$ in the right shoulder represent the intra- and inter-band contributions, respectively. In addition, to simplify the notation of the normal and anomalous Green functions, we abbreviate the parameters as follows.
\begin{eqnarray}
(-\bm{k}+\bm{q},i\omega_m-i\omega_n)\rightarrow(1),&\quad&(\bm{k},-i\omega_m)\rightarrow(2)\nonumber\\
(\bm{k}-\bm{q},i\omega_m+i\omega_n)\rightarrow(3),&\quad&(\bm{k},i\omega_m)\rightarrow(4)\nonumber
\end{eqnarray}
\begin{itemize}
\item Case with $\mu=\nu=a$\par
In this case, $\Phi_{aa}$ can be divided into three intra-band contributions labeled by $|j_a|$.
\begin{eqnarray}
\label{Phiaa}
&&\Phi_{aa}(\bm{q},i\omega_n)=\sum_{|j_a|=\frac{1}{2},\frac{3}{2},\frac{5}{2}}\Phi_{aa}^{\{|j_a|\}}(\bm{q},i\omega_n)\\
&&\Phi_{aa}^{\{|j_a|\}}(\bm{q},i\omega_n)=\frac{g^2\mu_B^2}{\beta}\sum_{\bm{k}}\sum_{\alpha,\beta=\pm|j_a|}(\hat{J}_a)_{\alpha\alpha}^2\left[F_{\alpha,\beta}^{\dagger}(1)F_{\alpha,\beta}(2)+G_{\alpha,\beta}(3)G_{\beta,\alpha}(4)\right]\quad\quad
\end{eqnarray}
\item Case with  $\mu=\nu=b$\par
In this case, $\Phi_{bb}$ can be divided into one intra- and four inter-band contributions,
\begin{eqnarray}
\label{Phibb}
&&\Phi_{bb}(\bm{q},i\omega_n)=\Phi_{bb}^{\{\frac{1}{2}\}}(\bm{q},i\omega_n)+\sum_{\{\alpha,\beta\}}\Phi_{bb}^{\{\alpha,\beta\}}(\bm{q},i\omega_n),
\end{eqnarray}
where, $\{\alpha,\beta\}=\{\frac{5}{2},\frac{3}{2}\},\{\frac{3}{2},\frac{1}{2}\},\{-\frac{1}{2},-\frac{3}{2}\},\{-\frac{3}{2},-\frac{5}{2}\}$, and each term is described as follows,
\begin{eqnarray}
&&\Phi_{bb}^{\{\frac{1}{2}\}}(\bm{q},i\omega_n)=\frac{g^2\mu_B^2}{\beta}\sum_{\bm{k}}\sum_{\alpha=\pm\frac{1}{2}}\left(\hat{J}_b\right)_{\frac{1}{2},-\frac{1}{2}}^2\Bigl[F_{\alpha,\alpha}^{\dagger}(1)F_{-\alpha,-\alpha}(2)+F_{\alpha,-\alpha}^{\dagger}(1)F_{-\alpha,\alpha}(2)\quad\quad\quad\nonumber\\
&&\quad\quad\quad\quad\quad\quad\quad\quad\quad\quad\quad\quad\quad\quad\quad+G_{\alpha\alpha}(3)G_{-\alpha,-\alpha}(4)+G_{\alpha,-\alpha}(3)G_{-\alpha,\alpha}(4)\Bigr],\quad\quad\quad\\
&&\Phi_{bb}^{\{\alpha,\beta\}}(\bm{q},i\omega_n)=\frac{g^2\mu_B^2}{\beta}\sum_{\bm{k}}\left(\hat{J}_b\right)_{\alpha,\beta}^2\Bigl[F_{\alpha,\alpha}^{\dagger}(1)F_{\beta,\beta}(2)+F_{\beta,\beta}^{\dagger}(1)F_{\alpha,\alpha}(2)\nonumber\\
&&\quad\quad\quad\quad\quad\quad\quad\quad\quad\quad\quad\quad\quad+F_{\alpha,-\alpha}^{\dagger}(1)F_{\beta,-\beta}(2)+F_{\beta,-\beta}^{\dagger}(1)F_{\alpha,-\alpha}(2)\nonumber\\
&&\quad\quad\quad\quad\quad\quad\quad\quad\quad\quad\quad\quad\quad+G_{\alpha,\alpha}(3)G_{\beta,\beta}(4)+G_{\beta,\beta}(3)G_{\alpha,\alpha}(4)\nonumber\\
&&\quad\quad\quad\quad\quad\quad\quad\quad\quad\quad\quad\quad\quad+G_{\alpha,-\alpha}(3)G_{-\beta,\beta}(4)+G_{\beta,-\beta}(3)G_{-\alpha,\alpha}(4)\Bigr].
\label{Phibb_alpha_beta}
\end{eqnarray}
\end{itemize}

\subsection{Spin susceptibility formulae}
The spin susceptibility formulae Eqs. (\ref{chiaa}) and (\ref{chibb}) are obtained by performing the analytical continuation and $\bm{q}$-limit on Eqs. (\ref{Phiaa}) and (\ref{Phibb}), respectively; i.e.
 $\Phi_{aa}^{\{|j_a|\}}(\bm{q},i\omega_n) \rightarrow \chi_{aa}^{\{|j_a|\}} $,  $\Phi_{bb}^{\{\frac{1}{2}\}}(\bm{q},i\omega_n) \rightarrow \chi_{bb}^{\{\frac{1}{2}\}} $, and 
$\Phi_{bb}^{\{\alpha,\beta\}}(\bm{q},i\omega_n) \rightarrow \chi_{bb}^{\{\alpha,\beta\}} $.
 In this section, we show the explicit forms of all the terms that appear on the right side of Eqs. (\ref{chiaa}) and (\ref{chibb}).  To simplify the notation, the following three new symbols are introduced,
\begin{eqnarray}
A_{j_a,\sigma}&=&\frac{1}{2E_{j_a\sigma}}\tanh{\frac{{\beta}E_{j_a,\sigma}}{2}},\\
B_{j_a,\sigma}&=&\frac{\beta}{4\left\{\cosh{\frac{{\beta}E_{j_a,\sigma}}{2}}\right\}^2},\\
C_{j_a,\sigma}&=&\frac{1}{E_{j_a,\sigma}^2}(B_{j_a,\sigma}-A_{j_a,\sigma}).
\end{eqnarray}
Here, the energy spectrum is given by $E_{j_a\sigma}=\sqrt{\varepsilon_{k,j_a}^2+d_0^2+|\bm{d}|^2+\sigma|\bm{q}|}$, and $d_0$ is the SC gap for the spin-singlet pairing. The detailed formulae are classified into four categories based on whether $d_0$, $\bm{d}$, and $\bm{q}$ are zero or not, respectively, as follows.
\begin{itemize}
\item Normal $\left(d_0=0,\,\bm{d}=0,\,\bm{q}=0\right)$
\begin{eqnarray}
&&\chi^{\{\alpha\}}_{aa}=g^2\mu_B^2\left(\hat{J}_a\right)_{\alpha,\alpha}^2\sum_{\bm{k}}2B_{\alpha,\sigma},\\
&&\chi^{\{\frac{1}{2}\}}_{bb}=g^2\mu_B^2\left(\hat{J}_b\right)_{\frac{1}{2},-\frac{1}{2}}^2\sum_{\bm{k}}2B_{\frac{1}{2},\sigma},\\
&&\chi^{\{\alpha,\beta\}}_{bb}+\chi^{\{-\beta,-\alpha\}}_{bb}=g^2\mu_B^2\left(\hat{J}_b\right)_{\alpha,\beta}^2\sum_{\bm{k}}\left[-4\frac{f\left(\varepsilon_{k,\alpha}\right)-f\left(\varepsilon_{k,\beta}\right)}{\varepsilon_{k,\alpha}-\varepsilon_{k,\beta}}\right].\quad(\alpha\ne\beta)
\end{eqnarray}

\item Spin-singlet $\left(d_0\ne0,\,\bm{d}=0,\,\bm{q}=0\right)$
\begin{eqnarray}
&&\chi^{\{\alpha\}}_{aa}=g^2\mu_B^2\left(\hat{J}_a\right)_{\alpha,\alpha}^2\sum_{\bm{k}}2B_{\alpha,\sigma},\\
&&\chi^{\{\frac{1}{2}\}}_{bb}=g^2\mu_B^2\left(\hat{J}_b\right)_{\frac{1}{2},-\frac{1}{2}}^2\sum_{\bm{k}}2B_{\frac{1}{2},\sigma},\\
&&\chi^{\{\alpha,\beta\}}_{bb}+\chi^{\{-\beta,-\alpha\}}_{bb}=g^2\mu_B^2\left(\hat{J}_b\right)_{\alpha,\beta}^2\sum_{\bm{k}}4\left[\frac{\varepsilon_{k\alpha}A_{\alpha,\sigma}-\varepsilon_{k\beta}A_{\beta,\sigma}}{\varepsilon_{k,\alpha}-\varepsilon_{k,\beta}}\right].\quad(\alpha\ne\beta)
\end{eqnarray}

\item Unitary $\left(d_0=0,\,\bm{d}\ne0,\,\bm{q}=0\right)$
\begin{eqnarray}
&&\chi^{\{\alpha\}}_{aa}=g^2\mu_B^2\left(\hat{J}_a\right)_{\alpha,\alpha}^2\sum_{\bm{k}}2\left[(\varepsilon_{k,\alpha}^2+|d_a|^2)C_{\alpha,\sigma}+A_{\alpha,\sigma}\right],\\
&&\chi^{\{\frac{1}{2}\}}_{bb}=g^2\mu_B^2\left(\hat{J}_b\right)_{\frac{1}{2},-\frac{1}{2}}^2\sum_{\bm{k}}2\left[(\varepsilon_{k,\frac{1}{2}}^2+|d_b|^2)C_{\frac{1}{2},\sigma}+A_{\frac{1}{2},\sigma}\right],\\
&&\chi^{\{\alpha,\beta\}}_{bb}+\chi^{\{-\beta,-\alpha\}}_{bb}=g^2\mu_B^2\left(\hat{J}_b\right)_{\alpha,\beta}^2\sum_{\bm{k}}4\left[\frac{\varepsilon_{k,\alpha}A_{\alpha,\sigma}-\varepsilon_{k,\beta}A_{\beta,\sigma}}{\varepsilon_{k,\alpha}-\varepsilon_{k,\beta}}\right].\quad(\alpha\ne\beta)
\end{eqnarray}

\item Nonunitary $\left(d_0=0,\,\bm{d}\ne0,\,\bm{q}\ne0\right)$
\begin{eqnarray}
&&\chi^{\{\alpha\}}_{aa}=g^2\mu_B^2\left(\hat{J}_a\right)_{\alpha,\alpha}^2\sum_{\bm{k},\sigma}\left[\varepsilon_{k,\alpha}^2\frac{q_a^2}{|\bm{q}|^2}C_{\alpha,\sigma}+A_{\alpha,\sigma}+\frac{2\sigma}{|\bm{q}|}\left(|d_a|^2+\varepsilon_{k,\alpha}^2\frac{q_b^2+q_c^2}{|\bm{q}|^2}\right)A_{\alpha,\sigma}\right],\nonumber\\\\
&&\chi^{\{\frac{1}{2}\}}_{bb}=g^2\mu_B^2\left(\hat{J}_b\right)_{\frac{1}{2},-\frac{1}{2}}^2\sum_{\bm{k},\sigma}\left[\varepsilon_{k,\frac{1}{2}}^2\frac{q_b^2}{|\bm{q}|^2}C_{\frac{1}{2},\sigma}+A_{\frac{1}{2},\sigma}+\frac{2\sigma}{|\bm{q}|}\left(|d_b|^2+\varepsilon_{k,\frac{1}{2}}^2\frac{q_a^2+q_c^2}{|\bm{q}|^2}\right)A_{\frac{1}{2},\sigma}\right],\nonumber\\\\
&&\chi^{\{\alpha,\beta\}}_{bb}+\chi^{\{-\beta,-\alpha\}}_{bb}=g^2\mu_B^2\left(\hat{J}_b\right)_{\alpha,\beta}^2\sum_{\bm{k},\sigma}2\left[\frac{\varepsilon_{k,\alpha}A_{\alpha,\sigma}-\varepsilon_{k,\beta}A_{\beta,\sigma}}{\varepsilon_{k,\alpha}-\varepsilon_{k,\beta}}\right].\quad(\alpha\ne\beta)
\end{eqnarray}
\end{itemize}

\subsection{Case with $\vec{H}~{\parallel}~a$-axis}
Here, we show the numerical results of the longitudinal spin susceptibility $\chi_{aa}$.
First, we show the results for unitary states in Fig. A$\cdot$1. The $d$-vectors for the unitary state are the same as in the case of $\chi_{bb}$, i.e. $\vec{d}=(k_a,k_b,k_c)$ and $\vec{d}=(0,k_c,k_b)$ ($A_u$ and $B_{3u}$ representations). In the case of $\vec{d}=(k_a,k_b,k_c)$, the decrease in $\chi_{aa}$ is about $30\sim35\%$, which is relatively larger than in the case of $\chi_{bb}$. This is because that all terms in Eq. (\ref{chiaa}) are intra-band contributions , and unlike in the case of $\chi_{bb}$, there are no inter-band terms that do not change in temperature. On the other hand, in the case of $\vec{d}=(0,k_c,k_b)$, the decrease is $0\%$. This is because that, in general, the spin susceptibility for fields perpendicular to the $d$-vector is unaffected by the superconducting transition.
\begin{figure}[H]
\begin{center}
\label{fig chiaa unitary}
\includegraphics[clip,width=13.5cm]{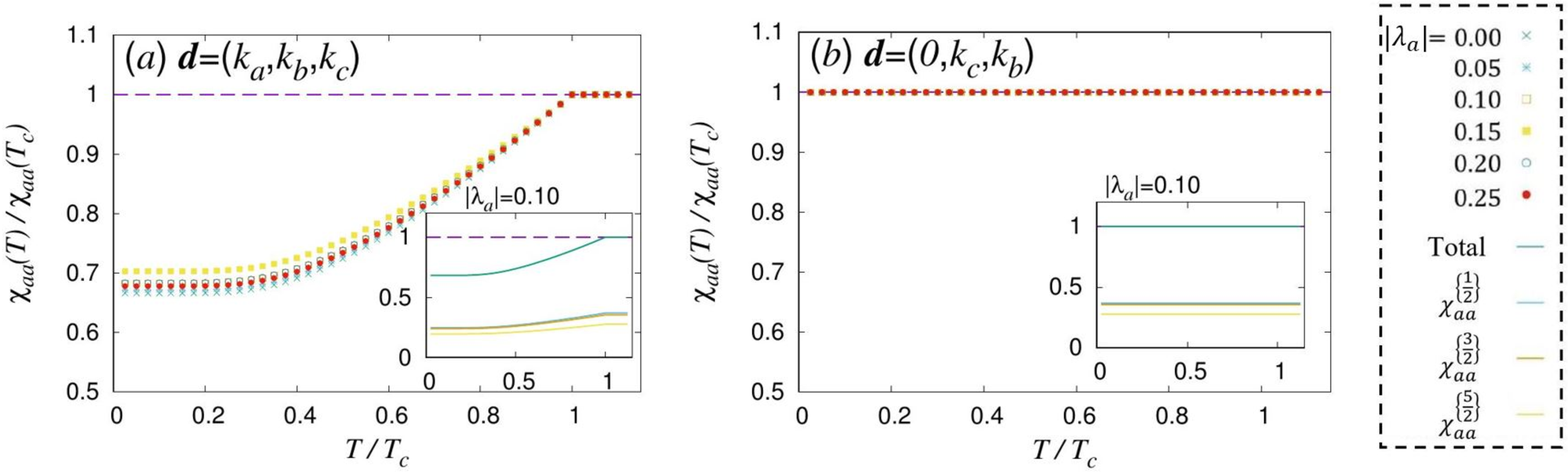}
\end{center}
\caption{(Color)The temperature dependence of $\chi_{aa}$ for the unitary state. (a)$A_u$ : $\vec{d}=(k_a,k_b,k_c)$. (b)$B_{3u}$ : $\vec{d}=(0,k_c,k_b)$.}
\end{figure}
Next, we show the results for non-unitary states in Fig. A$\cdot$2. 
\begin{figure}[b]
\begin{center}
\label{fig chiaa non-unitary}
\includegraphics[clip,width=13.5cm]{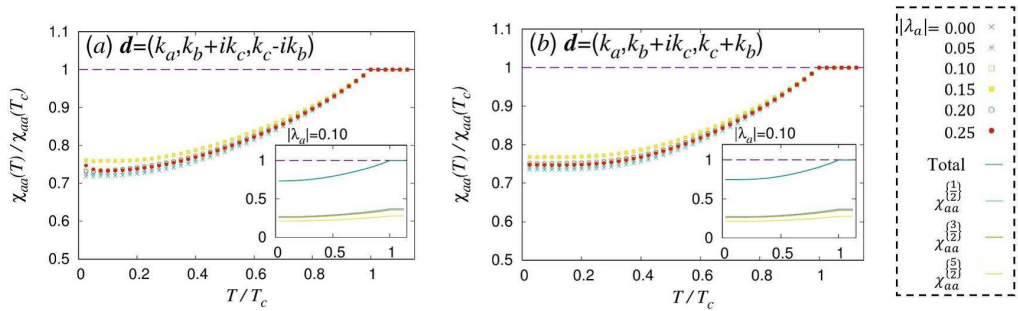}
\end{center}
\caption{(Color)The temperature dependence of $\chi_{aa}$ for the non-unitary state. (a) $A_{u}+B_{3u}$ : $\vec{d}=(k_a,k_b+ik_c,k_c-ik_b)$. (b) $A_{u}+B_{3u}$ : $\vec{d}=(k_a,k_b+ik_c,k_c+k_b)$.}
\end{figure}
The non-unitary states are considered to be realized when a strong magnetic field is applied in the $a$-axis direction,  then the Cooper pair is expected to carry the magnetization in the $a$-axis. This condition is represented by Eq. (\ref{qb}) for $\mu=a$. The followings are examples of $d$-vectors for non-unitary states belonging to the $A_{u}+B_{3u}$ representation.
\begin{eqnarray}
\vec{d}^{A_{u}+B_{3u}}_1&=&k_a\hat{a}+\left(k_b+ik_c\right)\hat{b}+\left(k_c-ik_b\right)\hat{c},\\
\vec{d}^{A_{u}+B_{3u}}_2&=&k_a\hat{a}+(k_b+ik_c)\hat{b}+\left(k_c+k_b\right)\hat{c}.
\end{eqnarray}
In the case of $\vec{d}^{A_{u}+B_{3u}}_1$, there is a line node at the crossing line of the $k_b$-$k_c$ plane and the Fermi surface. In the case of $\vec{d}^{A_{u}+B_{3u}}_2$, there are point nodes at the crossing points of the $k_c$-axis and the Fermi surface, and these are Weyl point nodes. The decreases in $\chi_{aa}$ are about $25\sim28\%$ in the case of $\vec{d}=(k_a,k_b+ik_c,k_c-ik_b)$, and about $26\sim27\%$ in the case of $\vec{d}=(k_a,k_b+ik_c,k_c+k_b)$. The decreases are relatively large compared to the case of $\chi_{bb}$. This is because that there are only intra-band contributions, as in the case of $\vec{d}=(k_a,k_b,k_c)$ in the unitary state described above.

\subsection{Case with the spin-singlet pairing}
In this section, we present the numerical results for $\chi_{aa}$ and $\chi_{bb}$ in the spin-singlet pairing case for comparison. The results are shown in  Figs. A$\cdot$3 and A$\cdot$4. 
The decrease in $\chi_{aa}$, in which there are only intra-band contributions, is $100\%$, as expected for conventional spin-singlet pairing states.  
On the other hand, even in the case of the spin-singlet case, the decrease in $\chi_{bb}$ is suppressed to $21\sim26\%$,
because of inter-band contributions which are not affected by the SC transition.
\begin{figure}[h]
\begin{center}
\includegraphics[clip,width=8cm]{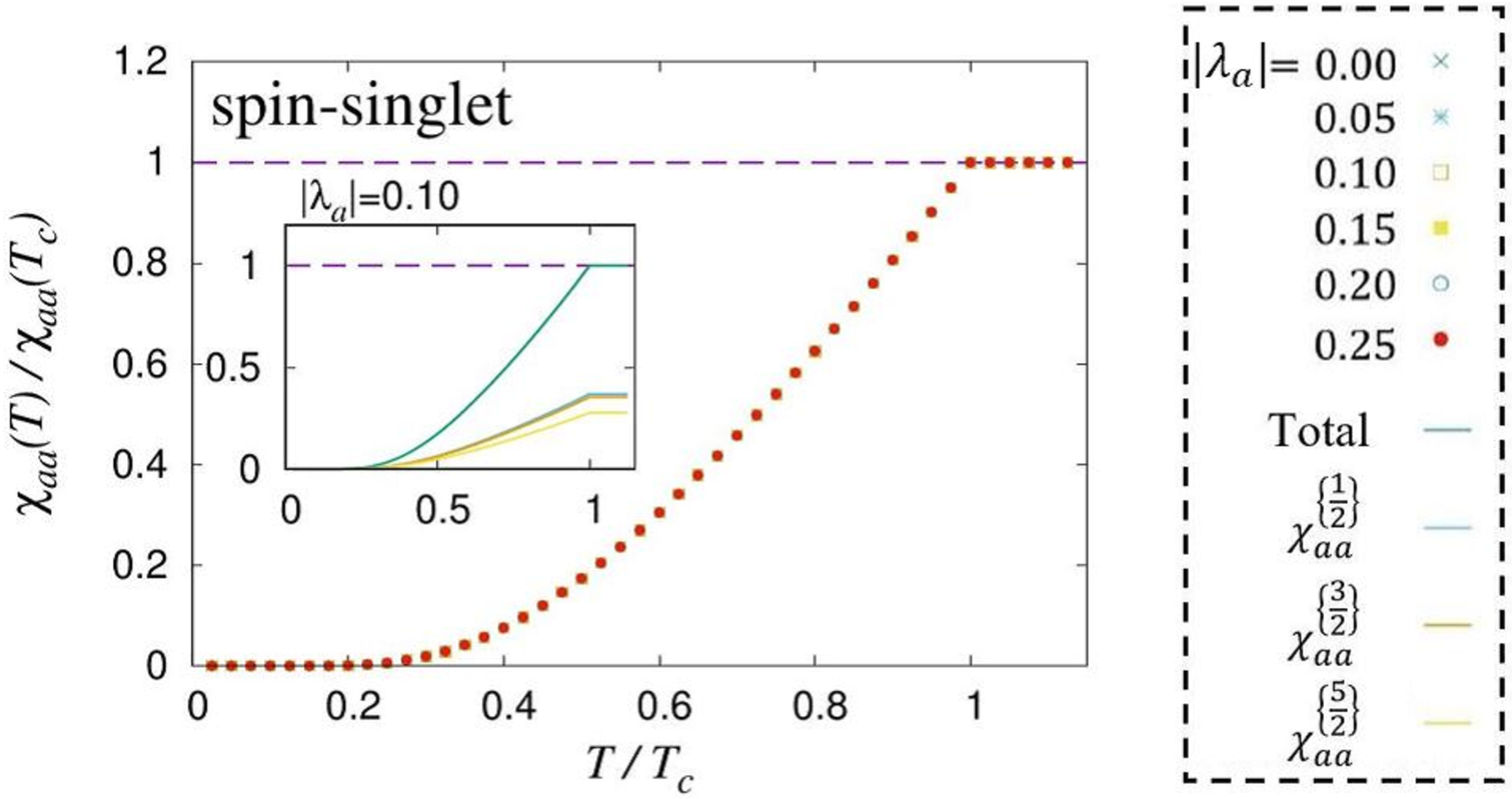}
\end{center}
\caption{(Color)The temperature dependence of $\chi_{aa}$ for the spin-siglet pairing}
\end{figure}

\begin{figure}[h]
\begin{center}
\includegraphics[clip,width=8cm]{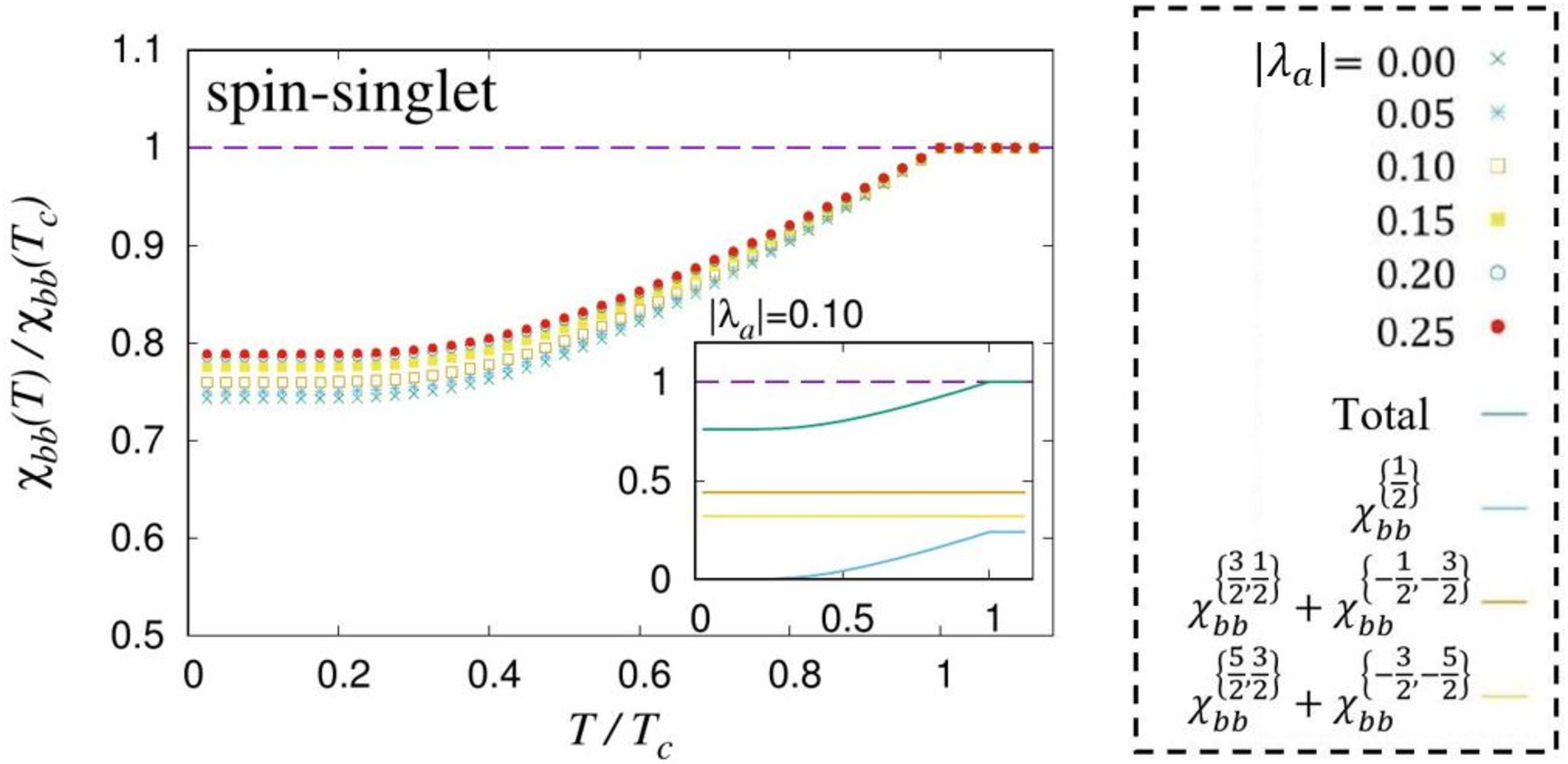}
\end{center}
\caption{(Color)The temperature dependence of $\chi_{bb}$ for the spin-siglet pairing}
\end{figure}


\end{document}